\documentstyle[12pt,epsfig]{article}                                
\begin{document}
\titlepage                  
\vspace{1.0in}
\begin{bf}
\begin{center}

Negative spectrum in Harmonic oscillator under simultaneous Non-Hermitian 
transformation of co-ordinate and momentum with Real wave function .
         
\end{center}
\end{bf}

\begin{center}
	      Biswanath Rath$\star$

Department of Physics, North Orissa University, Takatpur, Baripada -757003, Odisha, INDIA :$\star$ E.mail:biswanathrath10@gmail.com

\end{center} 
\vspace{0.5in}

We notice that $\mathcal{PT}$ symmetric non-Hermitian one dimensional simple 
Harmonic Oscillator under simultaneous transformation of co-ordinate and 
momentum with proper choice of positive oscillating frequency can reflect 
negative spectrum with well behaved wave function in real space .We also present a suitable comuter programme to realise the negative spectrum directly .

PACS(2008) :   03.65.Db

\vspace{0.2in}

Key words- Non-Hermitian Harmonic oscillator, Perturbation theory,well behaved
  Wavefunction ,real space ,positive frequency of oscillation ,
,direct study using comuter programming,netative spectrum.

\vspace{0.2in}
\begin{bf}
 
I.INTRODUCTION 
\vspace{0.2in}
\end{bf} 

In Quantum Physics,understanding on  two important things [1] :(i) well behaved wave function 
\begin{equation}
\Psi_{n}(x \rightarrow \infty ) \Rightarrow  0
\end{equation}

and (ii) commutation relation 
\begin{equation}
[x,p] = i 
\end{equation}
is necessary .Here $\Psi_{n}$ is the nth state wave function of the Hamiltonian
\begin{equation}
H |\Psi_{n} > = E_{n} | \Psi_{n}>
\end{equation}
having energy eigenvalue $E_{n}$  .The best example to understand wave function is to visualise  the problem of Harmonic Oscillator ,which is an interesting 
system in Quantum Physics .Its eigenvalue relation 
\begin{equation}
H_{HO}|\psi_{n}> =\frac{p^{2}}{2}+\frac{x^{2}}{2} |\psi_{n}>
\end{equation}
is well understood . Energy eigenvalue is 
\begin{equation}
E_{n}=(n+\frac{1}{2})  
\end{equation}

and corresponding wavefunction is 
\begin{equation}
\psi_{n}=\sqrt{\frac{1}{\sqrt{\pi}2^{n}n!}}H_{n}(x)e^{-\frac{x^{2}}{2}} 
\end{equation}
 where $H_{n}(x)$ is the Hermite polynomeal.
However we still believe that this simple  oscillator has not been studied 
in detail particularly under invariance in commutation relation.Let us discuss
the transformation 
\begin{equation}
x = ix
\end{equation}
prposed by Bender,Hook and Klevansky [2] . The Hamiltonian can be written as 
\begin{equation}
H(x\rightarrow ix) = -[\frac{p^{2}}{2}+\frac{x^{2}}{2}] 
\end{equation}
 whose energy eigenvalue is 
\begin{equation}
E_{n}= -(n+\frac{1}{2})  
\end{equation}
However the wave function represents 
\begin{equation}
\psi_{n}=\sqrt{\frac{1}{\sqrt{\pi}2^{n}n!}}H_{n}(x)e^{\frac{x^{2}}{2}} 
\end{equation}
divergent nature in real space i.e
\begin{equation}
\Psi_{n}(x \rightarrow \infty ) \Rightarrow  \infty
\end{equation}
Hence we believe that approach of BHK[2] is not a suitable representation in 
real space  as one can not visualise the negative energy in real space in view 
of divergent nature of wave function .In this context we notice that Fernandez[3] and Rath [4] independently discussed negative energy of Harmonic Oscillator 
under simultaneous transformation of co-ordinate  $x$ as [5] 
\begin{equation}
x \rightarrow \frac {x+i\lambda p}{\sqrt{(1+\beta\lambda)}} 
\end{equation}

and momentum $p$ as [5] 
\begin{equation}
 p \rightarrow \frac {p+i\beta x}{\sqrt{(1+\beta\lambda)}} 
\end{equation}
.It is worth to notice that both the authors have different opinion on 
visualising the negative spectrum .Rath [4] argues  with positive frequency of
oscillation $ w_{1} or w_{2}  \gg 0$ ,where as Fernandez [3] with negative frequency of 
oscillation $\omega \ll 0$ having divergent nature of wave function
\begin{equation}
 \psi_{0}(x) = <x|\psi_{0}> = \frac{\omega^{1/4}}{\pi^{1/4}}e^{-\omega x^{2}/2} 
\end{equation}
There fore one is likely to  believe that prposals of BHK[2] and Fernandez [3] even if suitable for complex space  but can hardly be a plausible approach for
 real space . Hence in this  paper we present a detail analysis of well behaved
 wave function in real space having negative energy of Harmonic Oscillator 
under simultaneous transformation of momentum and co-ordinate .Further we also 
present a direct simple approach to realise the  negative energy based on 
computational work .

\begin{bf}

II. Invariance in commutation relation and transformed Hamiltonian in second
 quantisation relation. 
\vspace{0.2in}
\end{bf}

Now we consider the commutation relation
\begin{equation}
      [ \frac{(x + i \lambda p)}{\sqrt{(1+\beta \lambda)}} , \frac{(p+i\beta x)}{\sqrt{(1+\lambda\beta)}} ] = [x,p] =i
\end{equation}
Now the new Hamiltonian with transformed $x$ and $p$ becomes non-Hermitian in nature and is 
\begin{equation}
 H= \frac {(p+i\beta x)^{2}}{2(1+\lambda\beta)}+\frac {(x+i\lambda p)^{2}}{2(1+\lambda\beta)} 
\end{equation}
In order to  solve  the above Hamiltonian (Eqn. (7)), we use the second quantization formalism as
\begin{equation}
x=\frac{(a+a^{+})}{\sqrt{2\omega}}    
\end{equation}

and
\begin{equation}
p=i\sqrt{ \frac{\omega}{2}} (a^{+}-a)    
\end{equation}

where the creation operator, $a^{+}$ and annhilation operator $a$ satisfy the commutation relation
\begin{equation}
[a,a^{+}]=1 
\end{equation}

and $\omega$ is an unknown parameter.
The transformed  Hamiltonian  can be written as 
\begin{equation}
H= H_{D} + H_{N} 
\end{equation}

where 
\begin{equation}
H_{D}=[(1-\lambda^{2})\omega+\frac{(1-\beta^{2})}{\omega}]\frac{(2a^{+}a+1)}{4(1+\lambda\beta)}   
\end{equation}

and 
\begin{equation}
H_{N}=U \frac{a^{2}}{4(1+\lambda\beta)} + V \frac{(a^{+})^{2}}{4(1+\lambda\beta)}     
\end{equation}

\begin{equation}
V=[-\omega (1-\lambda^{2})+\frac{(1-\beta^{2})}{\omega}-2(\lambda+\beta)]    
\end{equation}

\begin{equation}
U= [-\omega (1-\lambda^{2})+\frac{(1-\beta^{2})}{\omega}+2(\lambda+\beta)]   
\end{equation}

\begin{bf}

III(a).Negative  energy calculation (Case Study for U=0) 
\end{bf} 

 Now we solve the the eigenvalue relation:

\begin{equation}
H \Psi_{n}(x) = \epsilon_{n} \Psi_{n}(x)  
\end{equation}

using perturbation theory as follows.
Here we express 
\begin{equation}
\epsilon_{n} = \epsilon_{n}^{(0)}+\sum_{m=1}^{k}     \epsilon_{n}^{(m)} 
\end{equation}

The zeroth order energy $\epsilon_{n}^{(0)}$ satisfies the the following eigenvalue relation 
\begin{equation}
H_{D}|\psi_{n}> = H_{D}|n>=\epsilon^{(0)}_{n} |n> 
\end{equation}

where $\psi_{n}^{(0)}$ is the zeroth order wave function and $\epsilon_{n}^{(m)}$ is the mth order perturbation correction.
\begin{equation}
\epsilon_{n}^{(0)}= \frac{(2n+1)}{4(1+\lambda\beta)}[(1-\lambda^{2})\omega+ \frac{(1-\beta^{2})}{\omega}]  
\end{equation}

and
\begin{equation}
\sum_{m=1}^{k} \epsilon_{n}^{(m)}=\epsilon_{n}^{(1)}+\epsilon_{n}^{(2)}+\epsilon_{n}^{(3)}+....... 
\end{equation}

The  energy correction terms will give zero contribution if the parameter is determined from non-diagonal terms of $H_{N}$ [5]

Let the coefficient of $a^{2}$ is zero [5] i.e.
\begin{equation}
U= [-\omega (1-\lambda^{2})+\frac{(1-\beta^{2})}{\omega}+2(\lambda+\beta)]=0 
\end{equation}

which leads to
\begin{equation}
\omega =\omega_{1}=\frac{(\beta-1)}{(1+\lambda)}   
\end{equation}

In this case,
\begin{equation}
\epsilon_{n}^{(0)}= - (n+\frac{1}{2})  
\end{equation}

Now the perturbation correction term is 
\begin{equation}
H_{N}= V \frac{(a^{+})^{2}}{4(1+\lambda\beta)}=-\frac{(\lambda + \beta)}{1+\lambda \beta} (a^{+})^{2}     
\end{equation}

In this case one will notice that
\begin{equation}
<n| H_{N}|n-2> = V \frac{\sqrt{n (n-1)}}{4(1+\lambda\beta)}     
\end{equation}

\begin{equation}
<n-2| H_{N}|n> = 0      
\end{equation}

Hence it is easy to note that all orders of energy corrections will be zero.
Let us consider explicitly corrections up to third order using standard 
perturbation series given in literature[6-8], which can be written as
\begin{equation}
\epsilon_{n}^{(1)}=  <\psi_{n}|H_{N}|\psi_{n}>=0   
\end{equation}

\nonumber
$\epsilon_{n}^{(2)}= \sum_{k\neq n} \frac{<\psi_{n}|H_{N}|\psi_{k}><\psi_{k}|H_{N}|\psi_{n}>}{(\epsilon^{(0)}_{n}-\epsilon^{(0)}_{k})}    $
\begin{equation}
 =  \frac{<\psi_{n}|H_{N}|\psi_{n+2}><\psi_{n+2}|H_{N}|\psi_{n}>}{(\epsilon^{(0)}_{n}-\epsilon^{(0)}_{n+2})}=0    
\end{equation}

\nonumber
$ \epsilon_{n}^{(3)}=\sum_{p,q} \frac{<\psi_{n}|H_{N}|\psi_{p}><\psi_{p}|H_{N}|\psi_{q}><\psi_{q}|H_{N}|\psi_{n}>}{(\epsilon^{(0)}_{n}-\epsilon^{(0)}_{p})(\epsilon^{(0)}_{n}-\epsilon^{(0)}_{q})}=0    $

\begin{equation}
\epsilon_{n}^{(3)}= \frac{<\psi_{n}|H_{N}|\psi_{n+2}><\psi_{n+2}|H_{N}|\psi_{n+4}><\psi_{n+4}|H_{N}|\psi_{n}>}{(\epsilon^{(0)}_{n}-\epsilon^{(0)}_{n+2})(\epsilon^{(0)}_{n}-\epsilon^{(0)}_{n+4})}=0    
\end{equation}

Here second order correction is zero due to $<\psi_{n}|H_{N}|\psi_{n+2}>=\delta_{n,n+4}$ and third order correction is zero due to $<\psi_{n+4}|H_{N}|\psi_{n}>=\delta_{n+4,n+2}$. 
Similarly one can notice all correction terms $\epsilon^{(m)}_{n}$ will be zero. Hence 
\begin{equation}
\epsilon_{n} = \epsilon^{(0)}_{n} = - (n+\frac{1}{2})
\end{equation}

\begin{equation}
|\psi_{n}>= (\frac{\sqrt{\omega_{1}}}{\sqrt{\pi}2^{n}n!})^{\frac{1}{2}} H_{n}(\sqrt{\omega_{1}}x)e^{-\omega_{1} \frac{x^{2}}{2}} 
\end{equation}

with 
\begin{equation}
<\psi_{n}|\psi_{n}>=1 
\end{equation}

\begin{bf}
III.(b). Coresponding Wavefunction using Perturbation Theory
\end{bf}

Here we find the wavefunction as
\nonumber

$ \Psi_{n}^{(k)}=|\psi_{n}> + f_{\lambda,\beta}\frac{\sqrt{(n+2)!}}{2\sqrt{n!}} |\psi_{n+2}> + $
\nonumber
$ (f_{\lambda,\beta})^{2} \frac{\sqrt{(n+4)!}}{8\sqrt{n!}} |\psi_{n+4}> +$      
\begin{equation}
 (f_{\lambda,\beta})^{3} \frac{\sqrt{(n+6)!}}{48\sqrt{n!}} |\psi_{n+6}> +......
\end{equation}

where $f_{\lambda,\beta}=-\frac{(\lambda+\beta)}{(1+\lambda\beta)}$. 
In its compact form one can write,
\begin{equation}
\Psi_{n}^{(k)}=\sum_{k=0} (-1)^{k} [\frac{(\lambda+\beta)}{(1+\lambda\beta)}]^{k} \sqrt{\frac{(n+2k)!}{n!}} |\psi_{n+2k}>_{\omega_{1}} 
\end{equation}

The normalization condition here can be written as[6-9]
\begin{equation}
<\psi_{n}|\Psi_{n}^{(k)}> =1 
\end{equation}

So also the eigenvalue relation
\begin{equation}
<\psi_{n}|H|\Psi_{n}^{(k)}> = \epsilon_{n} = -(n+\frac{1}{2}) 
\end{equation}

\begin{bf}

III.(c). Direct calculation of negative energy using MATLAB 
\hspace{0.01in}
\end{bf}
Here we  would like to present a simple MATLAB programme [10,11] to demonstate
directly with a view to show that above selection of positive frequency can lead directly negative energy .
\begin{equation}
   \hspace{1.cm} N = 100 ; s = 1; 
\end{equation}
\begin{equation}
   \hspace{1.cm} n=1:N-1 ; 
\end{equation}
\begin{equation}
   \hspace{1.cm} m = \sqrt(n) ; 
\end{equation}
\begin{equation}
   \hspace{1.cm} \beta = 2  ; 
\end{equation}
\begin{equation}
   \hspace{1.cm} \lambda = 0.5  ; 
\end{equation}
\begin{equation}
   \hspace{1.cm}  L = \frac{1}{1 + \beta \lambda } ; 
\end{equation}
\begin{equation}
  \hspace{1.cm}  w_{1}= \frac{\beta -1}{\lambda + 1}  ; 
\end{equation}
\begin{equation}
    x = \frac{s}{\sqrt(2w_{1})} * (diag(m,-1) + diag(m,1) ; 
\end{equation}
\begin{equation}
    p = \frac{i\sqrt(w_{1})}{s\sqrt(2)} * (diag(m,-1) - diag(m,1) ; 
\end{equation}
\begin{equation}
   \hspace{1.cm}  R = p + \i \beta x ; 
\end{equation}
\begin{equation}
  \hspace{1.cm}  Y  = x + i \lambda p ; 
\end{equation}
\begin{equation}
  \hspace{1.cm}  H = 0.5*L (R^{2} + Y^{2} ) ; 
\end{equation}
\begin{equation}
  \hspace{1.cm}  Eig = sort(eig(H))     ; 
\end{equation}
\begin{equation}
 \hspace{1.cm}  Eig(1:100)    ; 
\end{equation}

In Table-1 , we present first fifty -0ne  eigenvalues of the transformed Harmonic Oscillator .

\bf {Table -I } : Negative Energy of transformed Harmonic Oscillator with positive frequency $ w_{1}=\frac{\beta-1}{\lambda+1}$ $\Longleftrightarrow  1,2,3,4$ \\ 
\begin{tabular}{|c|c|c|c|c|}  \hline
 n  & $\lambda=1,\beta=3$ & $\lambda=2,\beta=7$ & $ \lambda=0.5 ,\beta=5.5$&$\lambda=0.8,\beta = 8.2 $\\ \hline
 0  &  - 0.5    &- 0.5    &- 0.5     & - 0.5    \\ 
 1  &   -1.5    & -1.5    & -1.5     & - 1.5    \\ 
 2  & -  2.5    &- 2.5    & -2.5     & - 2.5    \\ 
 3  &  - 3.5    & -3.5    & -3.5     & - 3.5    \\ 
 4  &   -4.5    & -4.5    &- 4.5     & - 4.5    \\ 
 5  &   -5.5    & -5.5    & -5.5     & - 5.5    \\ 
 10  &  - 10.5    & -10.5    & -10.5     &  -10.5    \\ 
 20  &   -20.5    & -20.5    & -20.5     & - 20.5    \\ 
 30  &   -30.5    &- 30.5    & -30.5     &  -30.5    \\ 
 40  &   -40.5    & -40.5    & -40.5     & - 40.5    \\ 
 50  & -  50.5    & -50.5    & -50.5     &  -50.5    \\ \hline

 \end{tabular}     

\vspace{0.2in}

Here we notice eigenvalues are  negative for different values of $\beta$ and $\lambda$ with positive frequency of vibration .

\begin{bf}

IV.(a). Negative energy  calculation  (Case Study for V=0)
\hspace{0.01in}
\end{bf}

Let the coefficient of $(a^{+}){2}$ is zero [5] i.e.
\begin{equation}
V= [-\omega (1-\lambda^{2})+\frac{(1-\beta^{2})}{\omega}-2(\lambda+\beta)]=0 
\end{equation}
which leads to
\begin{equation}
\omega=\omega_{2} =\frac{(1+\beta)}{(\lambda -1 )}   
\end{equation}
Now the perturbation term becomes
\begin{equation}
H_{N}= U \frac{a^{2}}{4(1+\lambda\beta)}=\frac{(\lambda + \beta)}{1+\lambda \beta} a^{2}       
\end{equation}
In this case one will notice that
\begin{equation}
<\phi_{n}| H_{N}|\phi_{n+2}> = U \frac{\sqrt{[(n+1)(n+2)]}}{4(1+\lambda\beta)}  
\end{equation}

\begin{equation}
<\phi_{n+2}| H_{N}|\phi_{n}> = 0      
\end{equation}
Hence we have
\begin{equation}
\epsilon_{n}^{(1)}=  <\phi_{n}|H_{N}|\phi_{n}>=0   
\end{equation}

\begin{equation}
\epsilon_{n}^{(2)}= \sum_{k\neq n} \frac{<\phi_{n}|H_{N}|\phi_{k}><\phi_{k}|H_{N}|\phi_{n}>}{(\epsilon^{(0)}_{n}-\epsilon^{(0)}_{k})}    
 =  \frac{<\phi_{n}|H_{N}|\phi_{n-2}><\phi_{n-2}|H_{N}|\phi_{n}>}{(\epsilon^{(0)}_{n}-\epsilon^{(0)}_{n-2})}=0    
\end{equation}

\begin{equation}
\epsilon_{n}^{(3)}=\sum_{p,q} \frac{<\phi_{n}|H_{N}|\phi_{p}><\phi_{p}|H_{N}|\phi_{q}><\phi_{q}|H_{N}|\phi_{n}>}{(\epsilon^{(0)}_{n}-\epsilon^{(0)}_{p})(\epsilon^{(0)}_{n}-\epsilon^{(0)}_{q})}=0    
\end{equation}
or
\begin{equation} 
\epsilon_{n}^{(3)}= \frac{<\phi_{n}|H_{N}|\phi_{n-2}><\phi_{n-2}|H_{N}|\phi_{n-4}><\phi_{n-4}|H_{N}|\phi_{n}>}{(\epsilon^{(0)}_{n}-\epsilon^{(0)}_{n-2})(\epsilon^{(0)}_{n}-\epsilon^{(0)}_{n-4})}=0    
\end{equation}

Here second order correction is zero due to $<\phi_{n}|H_{N}|\phi_{n-2}>=\delta_{n,n-2}$ and third order correction is zero due to $<\phi_{n-4}|H_{N}|\phi_{n}>=\delta_{n-2,n-4}$. Similarly one can notice all correction terms $\epsilon^{(m)}_{n}$ will be zero. Hence 
\begin{equation}
\epsilon_{n} = \epsilon^{(0)}_{n}=E^{(0)}_{n}=-(n+\frac{1}{2})
\end{equation}
and
\begin{equation}
|\phi_{n}>= (\frac{\sqrt{\omega_{2}}}{\sqrt{\pi}2^{n}n!})^{\frac{1}{2}} H_{n}(\sqrt{\omega_{2}}x)e^{-\omega_{2} \frac{x^{2}}{2}} 
\end{equation}

\begin{bf}

IV.(b). Coresponding Wavefunction using Perturbation Theory
\end{bf}
\vspace{.5in}

Here we consider the wavefunction as

\vspace{.5in}

\nonumber
$ \Phi_{n}^{(k)}=|\phi_{n}> + f_{\lambda,\beta}\frac{\sqrt{n!}}{2\sqrt{(n-2)!}} |\phi_{n-2}> + 
 (f_{\lambda,\beta})^{2} \frac{\sqrt{n!}}{8\sqrt{(n-4)!}} |\phi_{n-4}> +  $
\begin{equation}
 (f_{\lambda,\beta})^{3} \frac{\sqrt{n!}}{48\sqrt{(n-6)!}} |\phi_{n-6}> +.....
\end{equation}
In its compact form, one can write
\begin{equation}
\Phi_{n}^{(k)}=\sum_{k=0}(-1)^{k}(\frac{(\lambda+\beta)}{(1+\lambda\beta)})^{k} \frac{\sqrt{n!}}{\sqrt{(n-2k)!}2^{k} k!}|\phi_{n-2k}>_{\omega_{2}} 
\end{equation}
Here we notice that for $x \rightarrow \infty $ i.e.
\begin{equation}
\phi_{n}(x \rightarrow \infty)\rightarrow 0 
\end{equation}
and
\begin{equation}
 \Phi_{n}^{(k)}(x \rightarrow \infty)\rightarrow 0 
\end{equation}
In this case, the normalization condition can be written as[6-9]
\begin{equation}
<\phi_{n}|\Phi_{n}^{(k)}> =1
\end{equation}
So also eigenvalue relation
\begin{equation}
<\phi_{n}|H|\Phi_{n}^{(k)}> =E_{n}=-(n+\frac{1}{2}) 
\end{equation}
\vspace{.2in}
\begin{bf}

IV.(c). MATLAB programme to realise negative energy directly.
\end{bf}
\vspace{.5in}
Here one has to use above MATLAB programme directly with new positive frequency of vibration $w_{2}$ as 
\begin{equation}
   \hspace{1.cm} \beta = 1  ; 
\end{equation}
\begin{equation}
   \hspace{1.cm} \lambda = 3  ; 
\end{equation}
\begin{equation}
  \hspace{1.cm}  w_{1}= \frac{\beta + 1}{\lambda - 1}  ; 
\end{equation}

In Table-II , we present first fifty -0ne  eigenvalues of the transformed Harmonic Oscillator .

\bf {Table -I } : Negative Energy of transformed Harmonic Oscillator with positive frequency $ w_{1}=\frac{\beta + 1}{\lambda - 1}$ $\Longleftrightarrow  1,2,3,4$ \\ 
\begin{tabular}{|c|c|c|c|c|}  \hline
 n  & $\lambda=3,\beta=1$ & $\lambda=2,\beta=1$ & $ \lambda=1.5 ,\beta=0.5$&$\lambda=1.5,\beta = 1 $\\ \hline
 0  &  - 0.5    &- 0.5    &- 0.5     & - 0.5    \\ 
 1  &   -1.5    & -1.5    & -1.5     & - 1.5    \\ 
 2  & -  2.5    &- 2.5    & -2.5     & - 2.5    \\ 
 3  &  - 3.5    & -3.5    & -3.5     & - 3.5    \\ 
 4  &   -4.5    & -4.5    &- 4.5     & - 4.5    \\ 
 5  &   -5.5    & -5.5    & -5.5     & - 5.5    \\ 
 10  &  - 10.5    & -10.5    & -10.5     &  -10.5    \\ 
 20  &   -20.5    & -20.5    & -20.5     & - 20.5    \\ 
 30  &   -30.5    &- 30.5    & -30.5     &  -30.5    \\ 
 40  &   -40.5    & -40.5    & -40.5     & - 40.5    \\ 
 50  & -  50.5    & -50.5    & -50.5     &  -50.5    \\ \hline

 \end{tabular}     

\vspace{0.2in}

Here we notice eigenvalues are  negative for different values of $\beta$ and $\lambda$ with positive frequency of vibration .

\begin{bf}
V. Conclusion
\end{bf} 

\vspace{.2in}

In this paper, we suggest a simpler procedure for calculating energy levels and wave function of the non-Hermitian harmonic oscillator under simultaneous
 transfromation of co-ordinate and momentum using perturbation theory .Further 
direct study on negative energy using simple MATLAB programme support the 
analytical study using perturbation theory .more interesting part is the 
well behaved nature of wave function under positive frequency of vibration i.e
\begin{equation}
   \Psi_{n}(w_{1} \gg 0, x\rightarrow \infty ) \rightarrow 0 
\end{equation}
\begin{equation}
   \Phi_{n}(w_{2} \gg 0, x\rightarrow \infty ) \rightarrow 0 
\end{equation}
which can hardly be seen in all previous work .


\begin{thebibliography} {99}    

\bibitem{Schiff} L.I.Schiff , Quantum Mechanics ,\bf{3rd Ed} ,Mc-Graw Hill Singapore (1985).
\bibitem{Bender} C.M.Bender,Daniel W.Hook and S.P.Klevansky,Negative energy$\mathcal{PT}$ symmetric Hamiltonians ,J.Phys.$\bf{ A 44}$,45(2012) ;arXiv:1203:0590v1[hep-th] .
\bibitem{Fernandez } F.M.Fernandez:Non-Hermitian Hamiltonian and similarity transformation.arXiv: 1502:02694v2[quantum-ph] .
\bibitem{Rath } B. Rath:Negative spectrum in Harmonic Oscillator under simultaneous Non-Hermitian transformation of co-ordinate and momentum with real wave-function. arXiv:1502.07891v1 [quantum-ph] .
\bibitem{Rath } B. Rath and P .Mallick .Zero-energy correction method for non-Hermitian harmonic Oscillator with simultaneous transformation of co-ordinate and
momentum :arXiv:1501.06161[quantum-ph].
\bibitem{Zettili} N.Zettili, Quantum Mechanics:Concepts and applications,{\bf 2nd ed}(John Wiley, New York, 2001),P-471.
\bibitem{Landau}  L.D. Landau and  E.M. Lifshtiz, Quantum Mechanics, {\bf 3rd ed}(Elsevier, Amsterdam, 2011).
\bibitem{Rath } B. Rath: A new approach on wave function and energy level calculation through perturbation theory. J. Phys. Soc. Jpn {\bf 67(9)}, 3044 (1998).
\bibitem{Rath } B. Rath: Case study of the convergency of nonlinear perturbation series: Morse-Feshbach nonlinear series. Int. J. Mod. Phys. {\bf A14(13)}, 2103 (1999).
\bibitem{Rath } B. Rath: Direct study on iso-spectral instability of general Harmonic Oscillator  under Non-Hermitian transformations : Breakdown of unbroken
Pseudo-Hermiticity and $\mathcal{PT}$ symmetry condition ,The African Review of Physics (2016) (in press) .
\bibitem{Korsch } H.J.Korsch and M.Gluck :Computing quantum eigenvalues made easy , Eur.J.Phys . {\bf 23}, 413 (2002)(This paper needs some mofication for direct study).
\end{thebibliography}
\end{document}